\documentclass[preprint,showpacs,preprintnumbers,amsmath,amssymb,nofootinbib,pre]{revtex4}

\usepackage{graphicx}
\usepackage{dcolumn}
\usepackage{bm}

\begin{document}

\preprint{JYFL}

\title{Finite-size effects in dynamics of zero-range processes}

\author{Janne Juntunen$^1$}
\email{janne.k.juntunen@jyu.fi}
\author{Otto Pulkkinen$^2$}
\email{otto.pulkkinen@lusi.uni-sb.de}
\author{Juha Merikoski$^1$}
\email{juha.t.merikoski@jyu.fi}

\affiliation{$^1$Deparment of Physics, University of Jyv\"askyl\"a\\
P.O.~Box 35, FI-40014 Jyv\"askyl\"a, Finland}

\affiliation{$^2$Fachrichtung Theoretische Physik\\ Universit\"at des Saarlandes\\ 
66123 Saarbr\"ucken, Germany}


\date{\today}

\begin{abstract}

The finite-size effects prominent in zero-range processes exhibiting a condensation transition are studied by using continuous-time Monte Carlo simulations. We observe that, well above the thermodynamic critical point, both static and dynamic properties display fluid-like behavior up to a density $\rho_c(L)$, which is the finite-size counterpart of the critical density $\rho_c = \rho_c(L\to\infty)$. We determine this density from the cross-over behavior of the average size of the largest cluster. We then show that several dynamical characteristics undergo a qualitative change at this density. In particular, the size distribution of the largest cluster at the moment of relocation, the persistence properties of the largest cluster and correlations in its motion are studied.

\end{abstract}



\maketitle

\section{\label{sec:Intro}Introduction}

Phase transitions in infinite systems appear as multiple solutions to the balance equations that derive from the generators for their time evolution. This multitude of possible stationary states is reflected by limit laws that in the thermodynamic limit fix the state with certainty: The phase space paths tend to fixed points determined by the system parameters.  Often a simple mathematical description in terms of deterministic differential equations is adequate. This is in contrast to {\it finite} physical or biological systems, which can escape the proximity of attractors again and again because of fluctuations due to stochasticity in microscopic processes so that the system is able to overcome free energy barriers. As an example, recent progress in imaging \cite{Yu06} has enabled study of cell phenotype due to stochastic transcriptional regulation of gene expression, which is typically controlled by a relatively small number of molecules in a bacterial cell \cite{Ozbudak04}. New studies not only show the effect of low copy number noise on the static properties of macroscopic variables, but also on their {\it dynamics} \cite{Mettetal06}. Models of gene expression often are (0+1)-dimensional and describe a 'well stirred cell'. The noise can play an equally important role in systems with non-zero spatial dimension, such as in pattern formation \cite{Cross93}.

The zero-range process (ZRP) is a paradigmatic model of spatially 
extended stochastic dynamics leading to a phase transition. 
It was first introduced as an example of an interacting 
Markov process in the 1970's by Spitzer~\cite{Spitzer70}. 
The fundamental difference between ZRP and most other condensation 
models is that many of its properties can be obtained analytically. 
In particular, Evans shows in Ref.~\cite{Evans00} that for 
condensation to happen on a regular graph in the thermodynamic limit (see also Ref.~\cite{Bialas97}), 
the transition rate function $u(k)$, where $k$ is the number of particles 
on a node, must approach its large $k$ value more slowly than $u(k)=1+2/k$. 
In~Ref.~\cite{Jeon00}, Jeon et al.~show that functions going to 
zero faster than $u(k)=\exp(-c\log^{\alpha}(k))$ 
induce a condensate in the system as well.
Also the effects of the topology of the underlying 
graphs on the dynamics have been considered \cite{Tang06}.
A recent comprehensive review is given by Evans in Ref.~\cite{Evans05}.

Although the stationary properties and the dynamics concerning formation of 
the condensate are well understood \cite{Evans00,Evans05}, the dynamics of the 
condensate after it has emerged has gained much less attention. 
In Ref.~\cite{Godreche03,Godreche05}, Godreche and Luck studied the stationary dynamics 
for the case $u(k)=1+b/k$ with $b>2$. 
In their analysis, they made the assumption that there are at most 
two primitive condensates simultaneously in the system, which is 
valid deep inside the condensed phase. 
In Ref.~\cite{Schwarzkopf08}, Schwarzkopf et al.~studied also dynamical 
properties, but in a model where multiple condensates were forced to emerge. 
For the static properties of ZRP, a detailed canonical analysis was
given by Evans et al.\ \cite{Evans06,Evans08}.
The finite-size effects, in particular the behavior of the current 
in the driven one-dimensional case, were studied by Gupta et al.\ \cite{Gupta07} 
and very recently, as interpreted via the concept of metastability, 
by Chleboun and Gro{\ss}kinsky \cite{Chleboun10} 
for $u(k)=1+b/k^\gamma$ with $\gamma < 1$.
These studies show that in general finite-size effects are 
prominent up to quite large system sizes.

In this paper, we use Monte Carlo simulations of symmetric one-dimensional 
ZRP with $u(k)=1+b/k$ to study the effect of the finite size $L$ of the 
system on quantities describing both static and some dynamical properties 
of ZRP close to the critical density of condensation. In particular, 
we shall consider the notion of the {\em fluid} and {\em condensed} 
phases and metastability (or coexistence) for finite systems. 
It turns out to be useful to define a finite-size counterpart 
$\rho_c(L)$ of the critical density $\rho_c = \rho_c(L\to\infty)$. 
We shall concentrate on the  behaviors of observables 
for $\rho_c < \rho < \rho_c(L)$.

\section{\label{sec:ZRP}ZRP and its basic properties}

The ZRP with particle conservation is a Markov process on a lattice or on a graph 
containing $L$ sites $i=1,2,\ldots,L$ and a fixed number $N$ of identical particles. 
The dynamics is defined via a transition rate function $u_i(k)$ for each site, 
i.e., a Markov rate, 
at which a site $i$ with $k$ particles on it looses a particle to other sites. 
The zero-range property means that the rate $u_i$ does not depend on the occupation 
numbers  of the other sites. The destination site for each particle move is 
determined by a (weighted and directed) graph.

The set of all occupation numbers $\{k_{i}\}$ defines the state of the system, and  
the probability that the system is in configuration $\lbrace k_i \rbrace$ in the stationary 
state is \cite{Spitzer70}
\begin{eqnarray}
 && P(\lbrace k_i \rbrace)=\frac{1}{\mathcal{Z}(L,N)}\prod^{L}_{i=1}f_{i}(k_{i}),
 \label{eq:Pm}\\
 && f_i(k) =
     \prod^{k}_{j=1} \frac{1}{u_i(j)} \mbox{ for $k\ge 1$ and }  
     f_i(0) = 1.
  \label{scal}
\end{eqnarray}
This result can easily be verified by noting that the 
probability distribution satisfies the local balance conditions for particle flows in and out of each site. 
The normalization factor 
\begin{equation}
 \mathcal{Z}(L,N)= \sum _{k_{1},k_{2},...,k_{L}}\delta\Bigl(\sum^{L}_{i=1}k_{i}-N\Bigr)\prod^{L}_{i=1}f_{i}(k_{i}).
 \label{Z}
\end{equation} 
is the partition function of the ZRP. 
In this paper, we consider the model with homogeneous rates 
\begin{equation}\label{eq:evans1}
  u_i(k) = u(k) = u_{0}(1+\frac{b}{k}),
\end{equation}
with constants $u_0$ and $b$, which was introduced in Refs.\ \cite{Evans00,Jeon00}. For $b>2$, this model has a continuous phase transition at the critical density 
\begin{equation}\label{density}
 \rho_{c}=\frac{1}{b-2},
\end{equation}
above which the system separates to a condensate consisting of a macroscopic number of particles 
\begin{equation}\label{eq:condensate1}
 Z_{1}=L(\rho-\rho_{c})  
\end{equation}
on a single, randomly located site, and to a homogeneous background elsewhere, which is described by a grand-canonical measure \cite{Grosskinsky03}. The statistics of the fluctuations of the condensate size depend on whether $2<b<3$ or $b>3$ \cite{Grosskinsky03,Majumdar05}.

We emphasize that the results above are obtained via a grand-canonical 
analysis, which is exact in the thermodynamic limit only, and finite 
systems can have significant deviations from them. 
In this paper, we report simulation results for the 
static and dynamic properties of the condensate 
in \emph{finite} one-dimensional ZRP with symmetric 
(non-driven) nearest-neighbor jumps and discuss them in light of 
existing analytical results \cite{Evans06,Evans08,Beltran08,Chleboun10}.

\newpage
\section{\label{sec:results}Results} 

\begin{figure}[h]
	\begin{center}
		\includegraphics[width=0.50\textwidth]{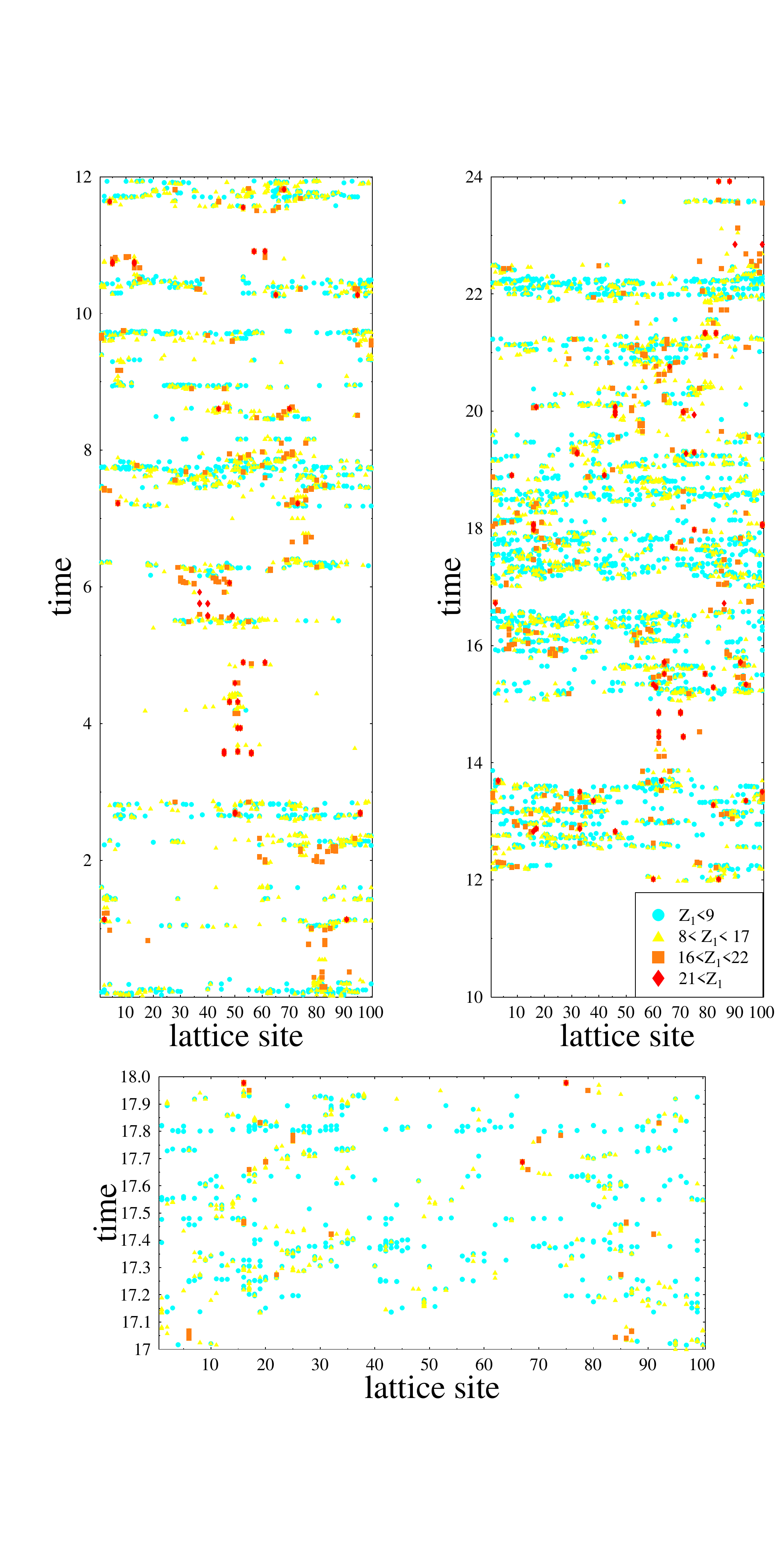}
	\end{center}
	\caption{(Color online) The location of the largest cluster as a function of time 
	for $b=5$, $\rho=3/4$ and $L=100$. Each symbol marks the position of the 
	largest cluster just after a jump and also 
	indicates the size of the cluster at that moment. In the lower panel 
	we show a magnification, from just below the middle of the upper right panel.
	The time axis has been scaled by a factor $10^5$.}
	\label{fig:configuration}
\end{figure}

The time evolution of the model of Eq.~(\ref{eq:evans1}) was simulated\footnote{The initial conditions for 
our simulations were completely random, after which the measured quantities had apparently converged 
in $10^7...10^9$ particle jumps. Even after the apparent equilibration of the dynamics (as seen e.g.~in the occupancy of the largest 
cluster) we required the largest cluster to relocate several thousand times before sampling the quantities 
of interest.} using a standard continuous-time Monte Carlo algorithm \cite{Bortz75}. 
As seen in a typical trajectory of the largest cluster, shown in Fig.~\ref{fig:configuration}, the dynamics of a finite system is bursty, in that the time evolution can be divided into two types of intervals: either little or no activity at all, or with rapid movement of the largest cluster. The same observation has been made in Fig.~3 of Ref.~\cite{Chleboun10} for a ZRP with interaction $u(k)=1+b/k^{\gamma}$, where $\gamma<1$, cf.~also Fig.~1 of Ref.~\cite{Godreche05}. Even if the switching between these 'phases' is perhaps not sharp for the interaction of Eq.~(\ref{eq:evans1}), as argued in Ref.~\cite{Chleboun10}, one can clearly discern the existence of several timescales in the time traces, from the length of duration of these intervals, down to the typical relocation time of the largest cluster and further to that of the motion of particles in the background. It is this multiple time-scale separation, which also makes the simulations very time-consuming in certain regions of the parameter space.

In Fig.~\ref{fig:configuration}, the colors indicate the size of the largest cluster just after a jump. Observe that the largest cluster tends to have a moderate size within the intervals of rapid movement, and that the largest observed values accumulate at the edges of these intervals. The latter cases involve a condensate. In the following section, we show how the simulation data can be used to determine an effective critical density for a finite system, but we investigate the size of the largest cluster at a non-specified time first.

\vfill\eject

\subsection{\label{sec:size}Size of the largest cluster} 

\begin{figure}[h]
	\begin{center}
		\includegraphics[width=0.60\textwidth]{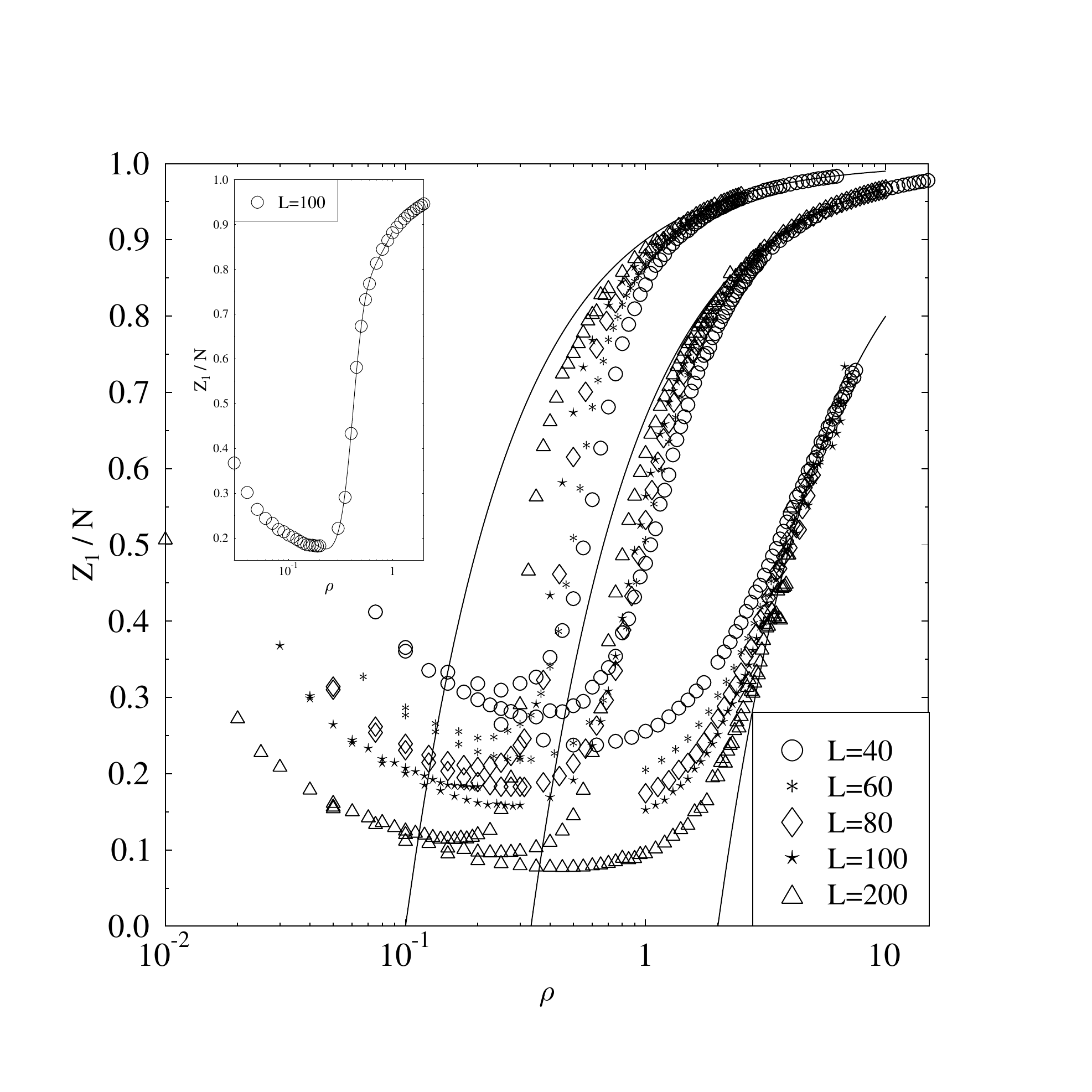}
	\end{center}
	\caption{The size of the largest cluster as a function 
	of the density for $b=2.5,5,12$ (from bottom to top). 
	The symbols denote the Monte Carlo data and  
	the full curves show the behavior at the thermodynamic 
	limit according to Eq.~(\ref{eq:anal}). 
	In the inset, we show a typical fit, here for $b=12$ and
	$L=100$, from which $\rho_c(L)$ is found at the 
	turning point of the fitted function.}
	\label{fig:turningpoint}
\end{figure}

Fig.~\ref{fig:turningpoint} depicts the simulation data for the size of the largest cluster in finite systems as a function of the particle density. All these results conform to the predictions of the grand-canonical theory,  
\begin{equation}
\label{eq:anal}
\frac{Z_{1}}{N}=1-\frac{\rho_c}{\rho} = 1- \frac{1}{(b-2)\rho} ,
\end{equation}
equivalent to Eq.~(\ref{eq:condensate1}), in the high density limit. At low densities, on the other hand, the data for each system size converge to a single curve irrespective of the interaction strength $b$ because the particles do not interact at all in a very dilute zero-range gas. For the high-density limit with constant $L$, 
we have $Z_1 \to N$. Due to discreteness, the low-density limit becomes $Z_1 = 1 = N$ and the curves are non-monotonic.

\begin{figure}[htbp]
	\begin{center}
  \vspace{30mm}
	\includegraphics[width=0.60\textwidth]{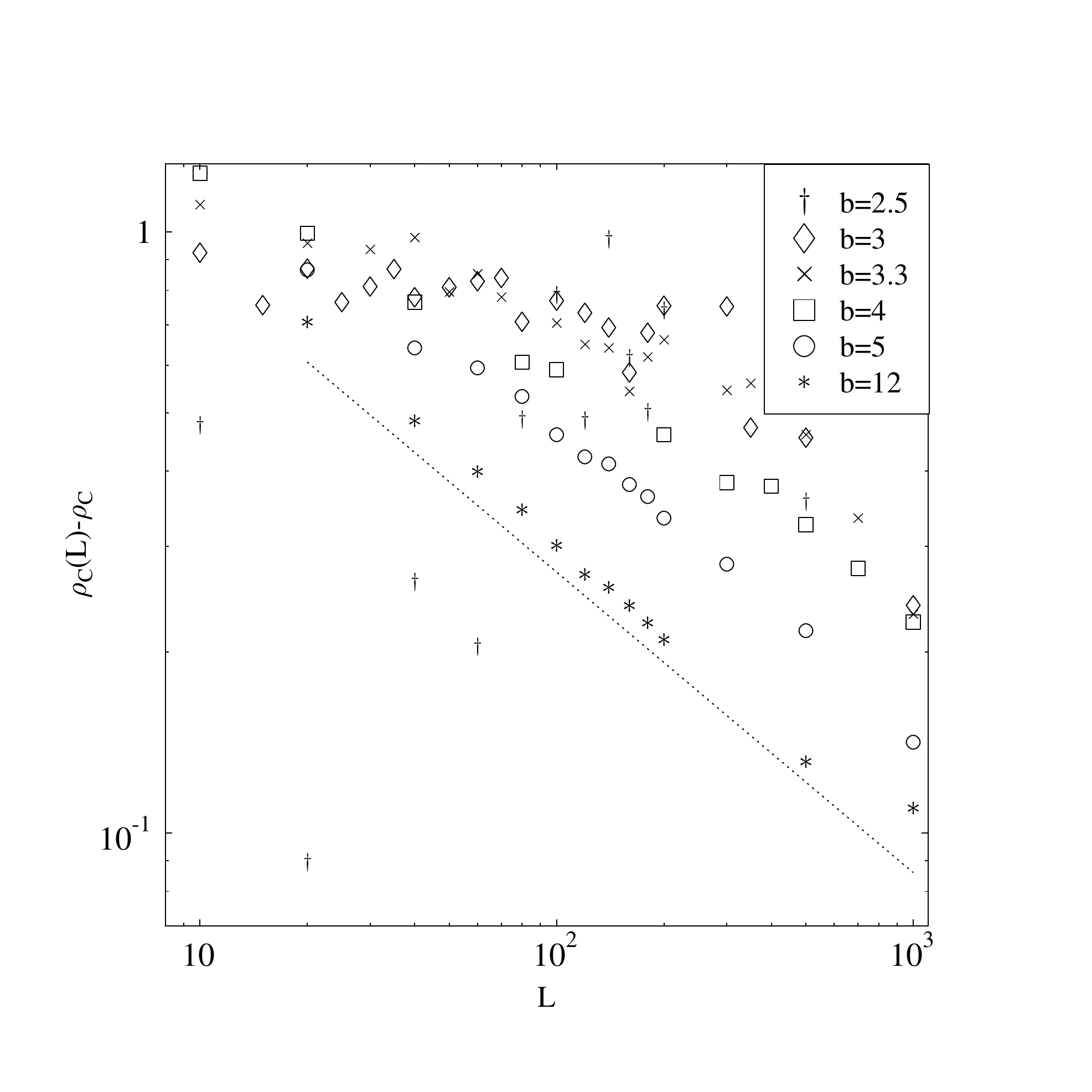}
	\end{center}
	\caption{Finite-size scaling of $\rho_c(L)$. 
	The slope of the dotted line corresponds to $\rho_c(L)-\rho_c \sim L^{-1/2}$. 
	Each data point has been determined by locating the turning point of 
	a curve $Z_1/N = Z_1(\rho)/\rho L$, cf.~Fig.~2.}
	\label{fig:scaling}
\end{figure}

We first determine the location of an effective transition point $\rho_c(L)$ by finding the turning point (zero of a second derivative) of the curves $Z_1/N = Z_1(\rho)/L\rho$. This method is an analog of locating the critical point of a second-order equilibrium phase transition from the finite-size 
scaling of the susceptibility peak. The results for $\rho_c(L)$ obtained this way 
are shown in Fig.~\ref{fig:scaling} for various system sizes and values of $b$. 

To set up a scaling hypothesis for $\rho_c (L)$, we note that, intuitively, 
one expects the collective features of a finite system to go through a smooth 
but rapid change, reminiscent of a phase transition in an infinite system, 
at a density marking the breakdown of scaling in distributions describing 
typical representatives of the system. One candidate for such a distribution 
is, in our case, the probability distribution for the cluster size on a site 
picked at random -- analogous, {\it e.g.}\, to a cluster size distribution in a 
percolation problem. Indeed, the mass distribution of a typical site in ZRP, 
derived in \cite{Evans06,Evans08}, shares the features of a distribution for 
the size of a randomly picked cluster in percolation \cite{Quinn76}: The 
distribution close to effective criticality exhibits a power law decay 
up to a  cutoff due to finite system size, and above this scale, 
an extra hump, describing respectively the mass in the condensate 
or a percolating cluster, emerges. In particular, the cluster size 
distribution of a typical site in a critical ZRP reads \cite{Evans06,Evans08}
\begin{equation}
\label{eq:pcrit2}
p(k)\sim  k^{-b}\exp(-k^{2}/2\Delta^2L).
\end{equation}
for $b>3$, and, for $2<b<3$, 
\begin{equation}
\label{eq:pcrit1}
p(k)\sim k^{-b} {V_{b}(k/L^{1/(b-1)})},
\end{equation}
where the scaling function $V_b$ decays to zero as a stretched exponential as the argument tends to infinity. 
These results imply that the cut-off density scales as $m_{\rm cutoff}/L \sim L^{-1/2}$ in the former case and 
as $L^{-(b-2)/(b-1)}$ in the latter case, respectively (see the argumentation in Sec.~4 of Ref.~\cite{Evans06}). 
We remark that the expected size of the largest cluster is in both cases of the order $L^{1/(b-1)}$, but this is not at variance with the $L^{1/2}$ scaling for $b>3$ because the scaling of the mean is a property of the power-law distribution below the cutoff, and the actual distribution for the size of the largest cluster also extends up to the scale $m_{\rm cutoff}$. For the same reason, we consider $m_{\rm cutoff}/L$ the more likely scale for $\rho_c (L)$ instead of the asymptotic scale of the mean   size of the largest cluster obtained from extremal statistics of independent variables without a cutoff.

The effective critical densities $\rho_c (L)$, plotted in Fig.~\ref{fig:scaling}, seem to conform to our hypothesis. We indeed observe the exponent $1/2$ for $b>3$. However, for $2<b<3$, the data becomes quite noisy (despite very long simulation times) and gives only a hint that the scaling could be different from the case $b>3$, i.e., of the form $L^{-(b-2)/(b-1)}$ as suggested by our hypothesis and Eq.~(\ref{eq:pcrit1}).  

\begin{figure}[h]
	\begin{center}
		\includegraphics[width=0.60\textwidth]{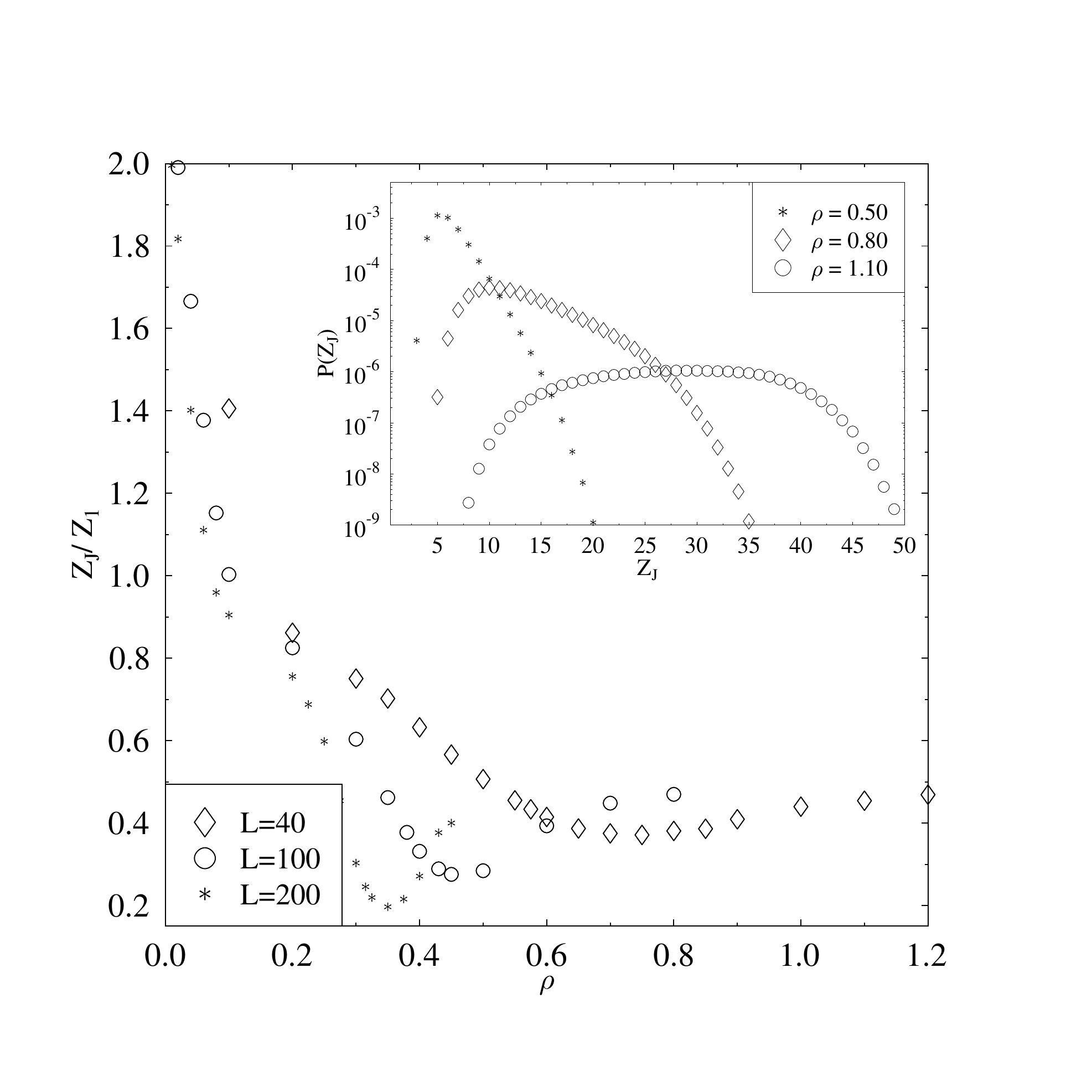}
	\end{center}
	\caption{The average size of the largest 
	cluster at the moment of relocation for $b=12$. 
	The inset shows the distribution of $Z_J$ for $b=5$ and $L=100$
	with the density smaller than, equal to, and greater than $\rho_c(L)\approx 0.8$
	obtained from Fig.~\ref{fig:scaling}.}
	\label{fig:jumpsize}
\end{figure}

In Fig.~\ref{fig:jumpsize}, we show data for another quantity, the statistics of which is expected to go through a change at the effective critical density $\rho_c(L)$, namely the size $Z_J$ of the largest cluster just after a jump of the condensate relative to size $Z_1$ of the largest cluster at a randomly chosen instant. There is a conspicuous dip in this ratio at intermediate densities. 
For the accessible system sizes (with $b>3$), the location of the dip is close to 
and scales (data not shown) the same way as $\rho_c(L)$.

The behavior of the ratio $Z_J/Z_1$ can be understood by considering the different statistical nature of $Z_J$ and $Z_1$ and the mechanisms of largest cluster relocation in different regimes. First, at low densities, the system is in the fluid phase, and the largest cluster changes its location frequently. The ratio $Z_J/Z_1$ is of the order one because the largest and the second largest cluster at a random instant are of the same order (e.g.~by an extremal statistics argument for grand-canonical measures), and a relocation event is a consequence of the two largest clusters being of the same size.{\footnote{The ratio $Z_J/Z_1$ exceeds unity in Fig.~\ref{fig:jumpsize} as $\rho\to 0$ because jumps of a largest cluster of size one are not considered a relocation in our simulation. For the relocation to have happened we require that the occupation of another site grows larger than that of the previous location.}} Second, at very high densities, nearly every particle belongs to the condensate, and a relocation occurs by splitting of the condensate into two equal-sized, transient condensates. Thus $Z_J/Z_1\to 1/2$ as $\rho\to \infty$. The dip in the ratio between these extremes then occurs because the system exhibits a mixture of fluid and condensate phases; the mean value of $Z_1$ is significantly larger than in the low density regime because of the great longevity of the condensate states. However, most of the counts to the statistics of $Z_J$ come from the fluid phase intervals with rapid largest cluster movement, and the sizes at the time of relocation in the fluid phase are of lower order than the size of a condensate.

The inset of Fig.~\ref{fig:jumpsize} shows the distributions of $Z_J$ at the three different regimes discussed above, and confirms our heuristic. The distribution is well fitted by an extremal statistics distribution at low densities, while its variance is large around the effective critical density $\rho_c(L)$. At high densities, the distribution gets peaked at large cluster sizes with its mean proportional to the size of the condensate.

\newpage
\subsection{\label{sec:lifetime}Largest cluster: persistence and correlation functions}

\begin{figure}[tbh]
	\begin{center}
		\includegraphics[width=0.60\textwidth]{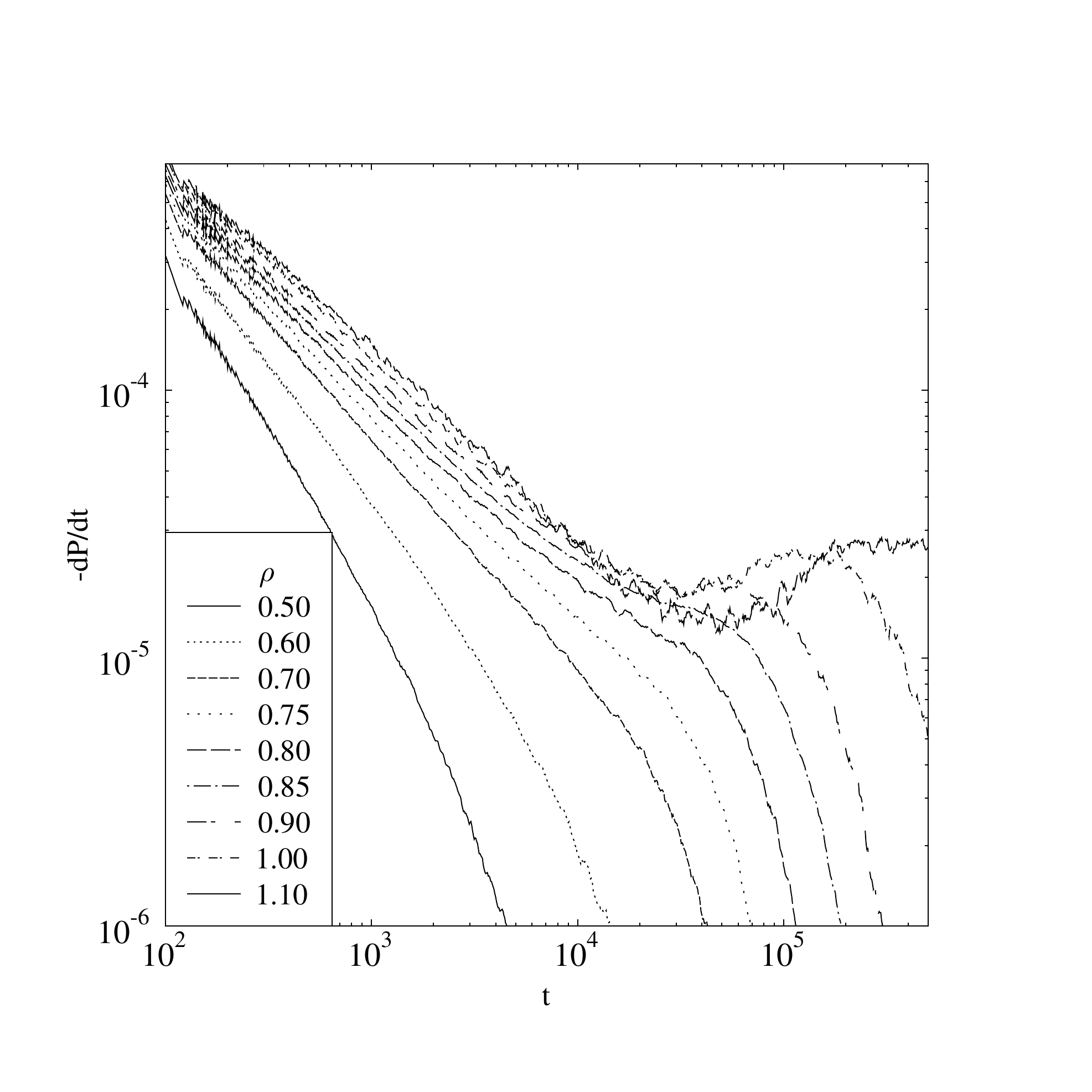}
	\end{center}
	\caption{The distribution of lifetimes of the largest cluster for $b=5$ and $L=100$ at various densities. 
	For this case $\rho_c(L) \approx 0.8$.}
	\label{fig:lifetime}
\end{figure}

The heuristic for the drop in the size of largest cluster at the time of relocation as compared to its usual average was based on the existence of fast and slow modes in the largest cluster movement. These are associated with the fluid and condensate phases, respectively. Fig.~\ref{fig:lifetime} shows that this is a valid assumption. The distribution of the lifetime of the largest cluster at a single site, i.e.~the derivative of the persistence probability $P(t)$ \cite{Majumdar96} that the largest cluster has not moved
between times $t_0$ and $t_0 +t$, is unimodal and has an exponential tail at densities smaller than $\rho_c(L)$. It changes to a power-law distribution with a cut-off due to finite size [cf.\ Eqs.~(\ref{eq:pcrit1}-\ref{eq:pcrit2})] at the transition density, and it eventually acquires a double-peak structure characteristic of the condensed phase \cite{Evans05,Chleboun10} at high densities. These observations are particularly interesting in the light of remark by Chleboun and Gro{\ss}kinsky \cite{Chleboun10} that, if one takes the number of transitions in the system as the order parameter, the model $u(k) = 1 +b/k$ is not metastable on the critical scale (both density and the number of transitions multiplied by $\sqrt{L/\log L}$). Nevertheless, metastability is present in the lifetime distributions without any scaling.

\begin{figure}[tbh]
	\begin{center}
	\includegraphics[width=0.60\textwidth]{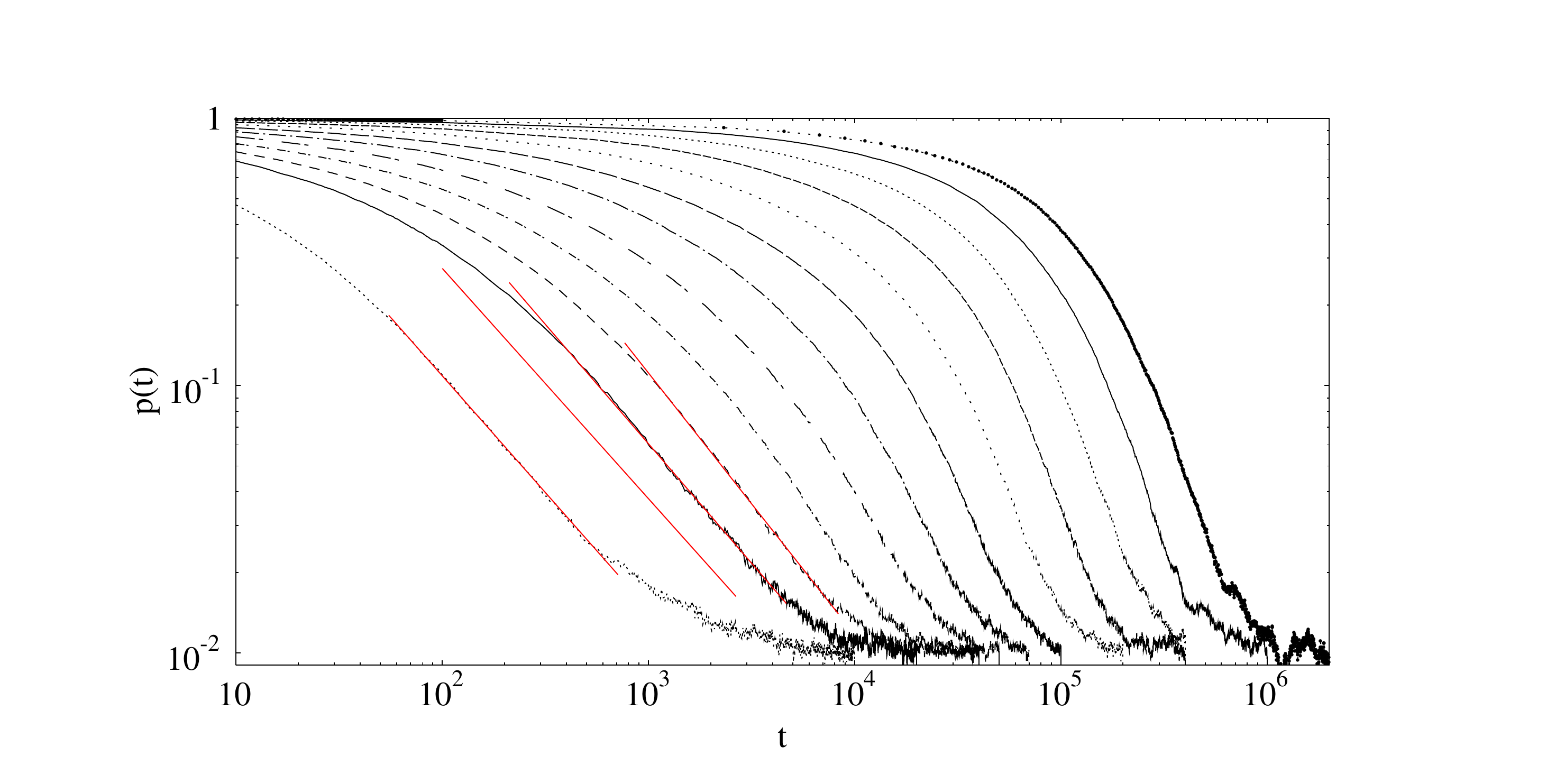}
	\includegraphics[width=0.60\textwidth]{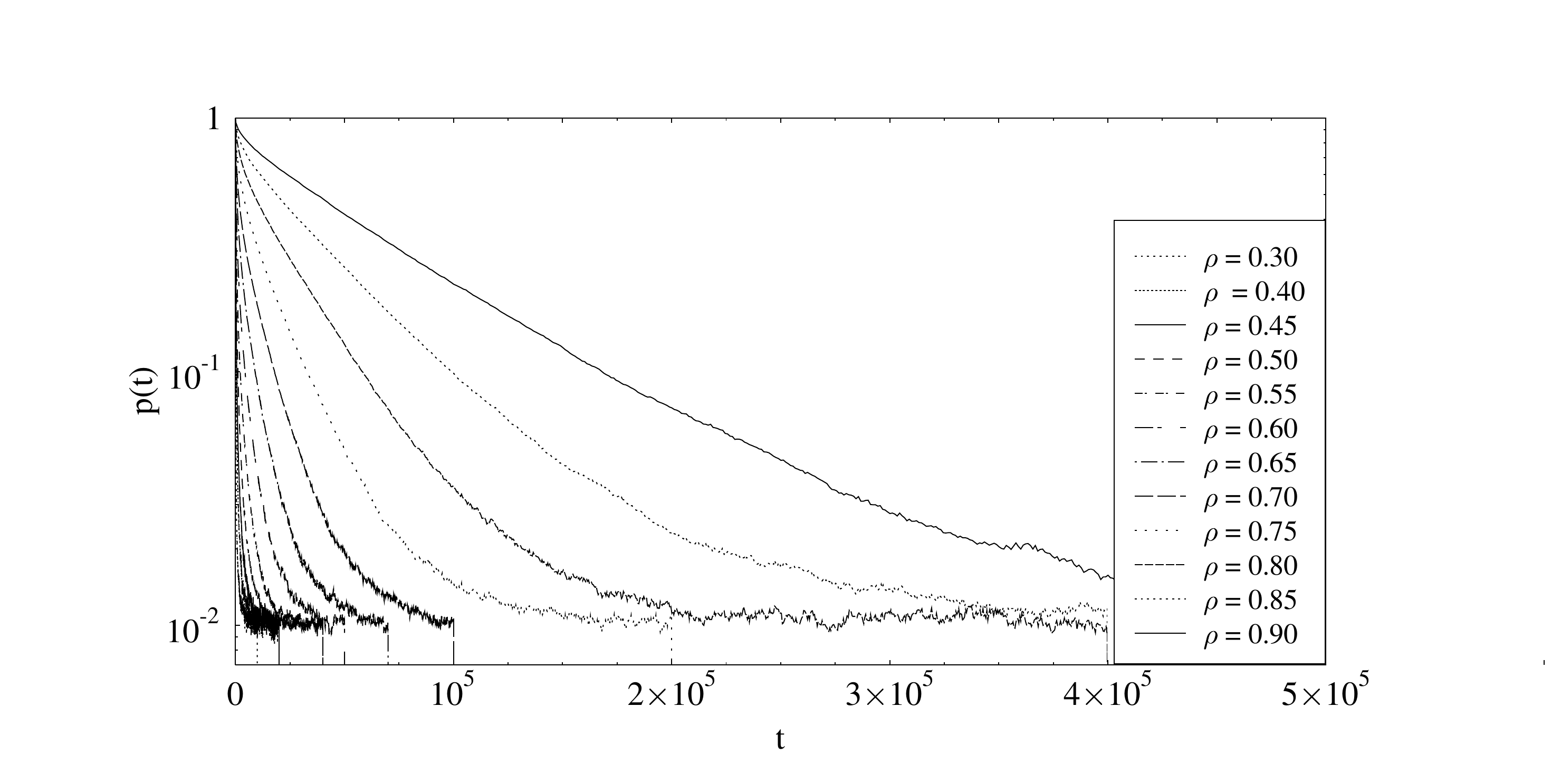}
	\end{center}
	\caption{The single-site correlation function as loglog and semilog plots 
	with fits to the apparent power law regions shown 
	for $b=5$ and $L=100$. The value of $\rho$ increases from bottom to top.}
	\label{fig:persistence}
\end{figure}

\begin{figure}[th]
	\begin{center}
	\includegraphics[width=0.60\textwidth]{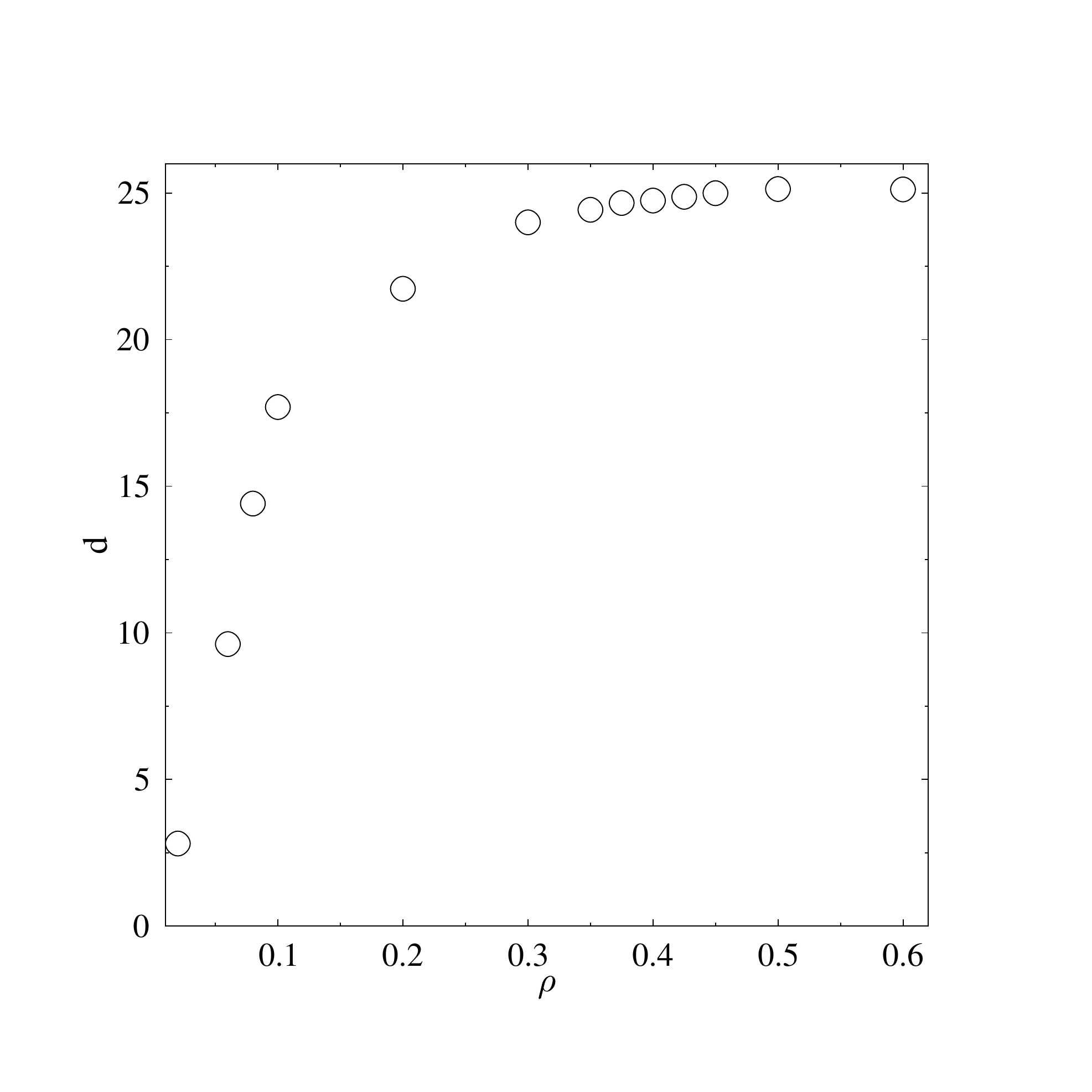}
	\end{center}
	\caption{The mean jump lengths for $b=12$ and $L=100$ at various densities. 
	For this case $\rho_c(L) \approx 0.4$.}
	\label{fig:spatial}
\end{figure}  
  
The lifetime distribution of the largest cluster tells how long the largest cluster stays on a single site, but contains no information on possible other correlations in occupation probabilities. For instance, it is quite likely that there is some extra mass left at a site recently occupied by the largest cluster, and this might introduce some memory in the relocation process. The time traces of Fig.~\ref{fig:configuration} support the existence of such mechanism. These correlations can be probed by measuring the single-site correlation function $p(t)= \langle I(X_{t_0} = i) I(X_{t_0+t}= i ) \rangle$, where $X_t$ is the position of the largest cluster at time $t$, and $I$ is the indicator function. This is the probability that the largest cluster occupies the same site at times $t_0$ and $t_0 +t$ irrespective of its motion in between these two times. 

Examples of correlation functions $p(t)$ are shown in Fig.~\ref{fig:persistence}. These are notable for two reasons. First, below $\rho_c(L)$, which in this case is around $\rho\approx 0.8$, we observe a power-law decay $p(t)\sim t^{-\alpha}$ for intermediate times. The exponents obtained from the fitted lines increase from roughly one to two as the density increases (however the range of the data available is too narrow to allow for a finite-size scaling analysis). The window of validity for the power-law fit at intermediate times shrinks to a point roughly at $\rho_c(L)$. Second, the semilogarithmic plot shows that the decay of the correlation at high densities is not exactly exponential, but more like a stretched exponential for a wide range of times.

The single-site correlation functions are fundamentally affected by three different physical mechanisms related to the relocation dynamics of the largest cluster, namely the persistence at a single site, the distribution of jump lengths, and the correlations in the jump lengths. All these measure how fast 'mixing' the stochastic movement of the largest cluster is. The importance of persistence is the most obvious because it distinguishes between certain and uncertain occupation of the site at a later time. That of the jump lengths can be made concrete by considering, for example, the return probabilities of simple random walks with different step lengths. Clearly, the longer the step, the more effectively the walker explores the state space, and the faster the correlations decay. The third mechanism, correlations in the jump lengths, is perhaps the most interesting of the three in spatially distributed systems, and is related to 'memory' induced by hidden degrees of freedom, such as transport of particles not belonging to the largest cluster. For example, the trajectory of the largest cluster can be, at least in principle, restricted to only a few sites for long periods of time because of some extra mass remaining at the recently occupied sites.

\begin{figure}[th]
	\begin{center}
	\includegraphics[width=0.60\textwidth]{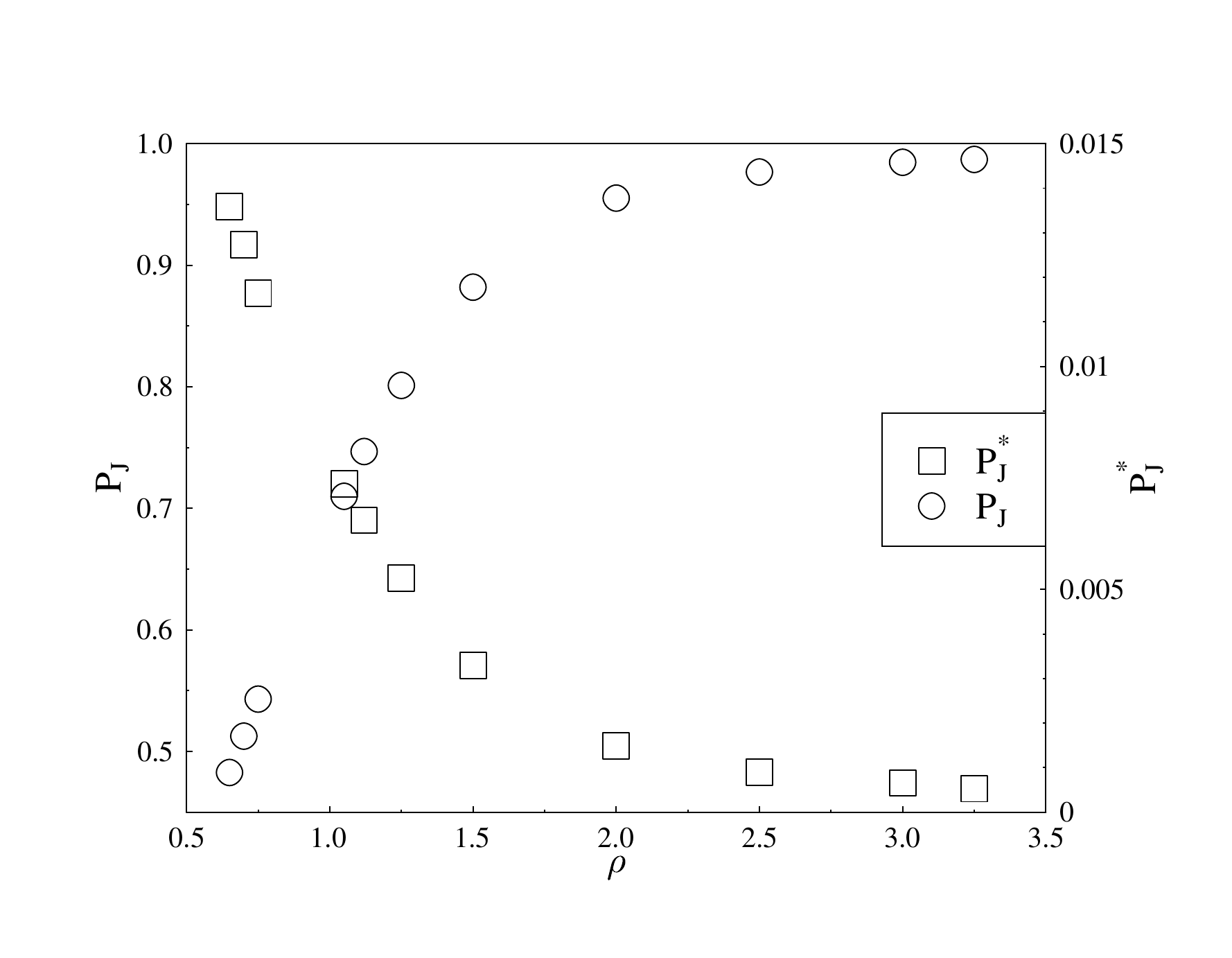}
	\end{center}
	\caption{Correlations in the jump lengths: 
	$P_{J}$ is the probability that two consecutive jumps of the largest cluster 
	have the same length 
	and $P_{J}^\ast$ is the same quantity with back-and-forth movement discarded.
	Here $b=5$ and $L=40$ so that $\rho_c(L)\approx 0.9$.}
	\label{fig:length_corr}
\end{figure}

Figure \ref{fig:spatial} excludes the jump length as the cause for the power-laws apparent in the single-site correlation functions because the jump lengths are of macroscopic order at the relevant densities. The lifetime distributions of Fig.\ \ref{fig:lifetime}, on the other hand, have a power-law part near the effective critical density, which can effect a slow decay of the correlation function. The proposed memory effect due to remnant mass on a recently occupied site is also present and can be very strong at large densities as is clear from Fig.\ \ref{fig:length_corr}. However, since we do not see that much broadening in the correlations at high densities, we mostly associate the power laws with the power laws in the lifetime distribution. The remnant mass could still be responsible for the stretched exponential decay of the correlation function.

We concluded that the mean jump lengths of the largest cluster are of macroscopic order at sufficiently high densities, and hence not a reason for the power-laws observed in the correlation functions. Indeed, it has been argued \cite{Godreche05} that in the thermodynamic limit, the condensate relocation is a completely random process. However, the jump lengths of the largest cluster can be small in subcritical systems, i.e., when $\rho<\rho_c (L)$. In Fig.~\ref{fig:spatial}, we show the average jump length for fixed $b$ even for very small densities in a finite system. In this example of size $L=100$, the average jump length for uncorrelated motion would be 25.5, which in this case is reached around $\rho_c(L)$. In the fluid phase, a diffusive mode is prominent. The largest cluster moves frequently in small steps, which is a consequence of low occupation numbers; the likelihood of close-by neighbors to have the same or nearly the same number of particles increases as the total particle density decreases, which, by movement of single particles, leads to short jumps of the largest cluster.

\newpage
\subsection{\label{sec:diffusion}Mass transport by diffusion} 

In addition to changes in the movement of the largest cluster, 
there are other dynamic indicators of the system switching between 
fluid and condensate phases. In this section, we briefly discuss 
the behavior of mass transport by diffusion in the framework 
of the Kubo-Green linear-response theory, where the
collective diffusion coefficient $D$ becomes the product of a
static response and the integral of the velocity autocorrelation
function or the center-of-mass diffusion coefficient $D_{cm}$.
This way $D$ can be written as
\begin{equation}
\label{eq:KG}
D(\rho) = \frac{1}{\rho\,K(\rho)}D_{cm}(\rho)
          \equiv f(\rho) D_{cm}(\rho),
\end{equation}
where both the thermodynamic contribution $f(\rho)$ and 
the mobility contribution $D_{cm}(\rho)$ depend on $\rho$. 
Here $K(\rho)$ is the compressibility, which is proportional to
the equilibrium density fluctuations, as seen in the
grand-canonical description:
$K \propto \langle (N- \langle N \rangle)^2\rangle/\langle N \rangle$,
which assumes Gaussian fluctuations around the most probable
particle density and strictly speaking would be modified for
our finite system with fixed $N$.
On the other hand, $D$ is also obtained in the standard way
from the decay of the Fourier transform of the density-density
autocorrelation function or the dynamic structure factor
(we note that the static structure factor could be used to
probe the possible existence of multiple condensates),
which is the solution of the corresponding diffusion
equation \cite{Chaikin95}.

\begin{figure}[htbp]
	\begin{center}
	\includegraphics[width=0.60\textwidth]{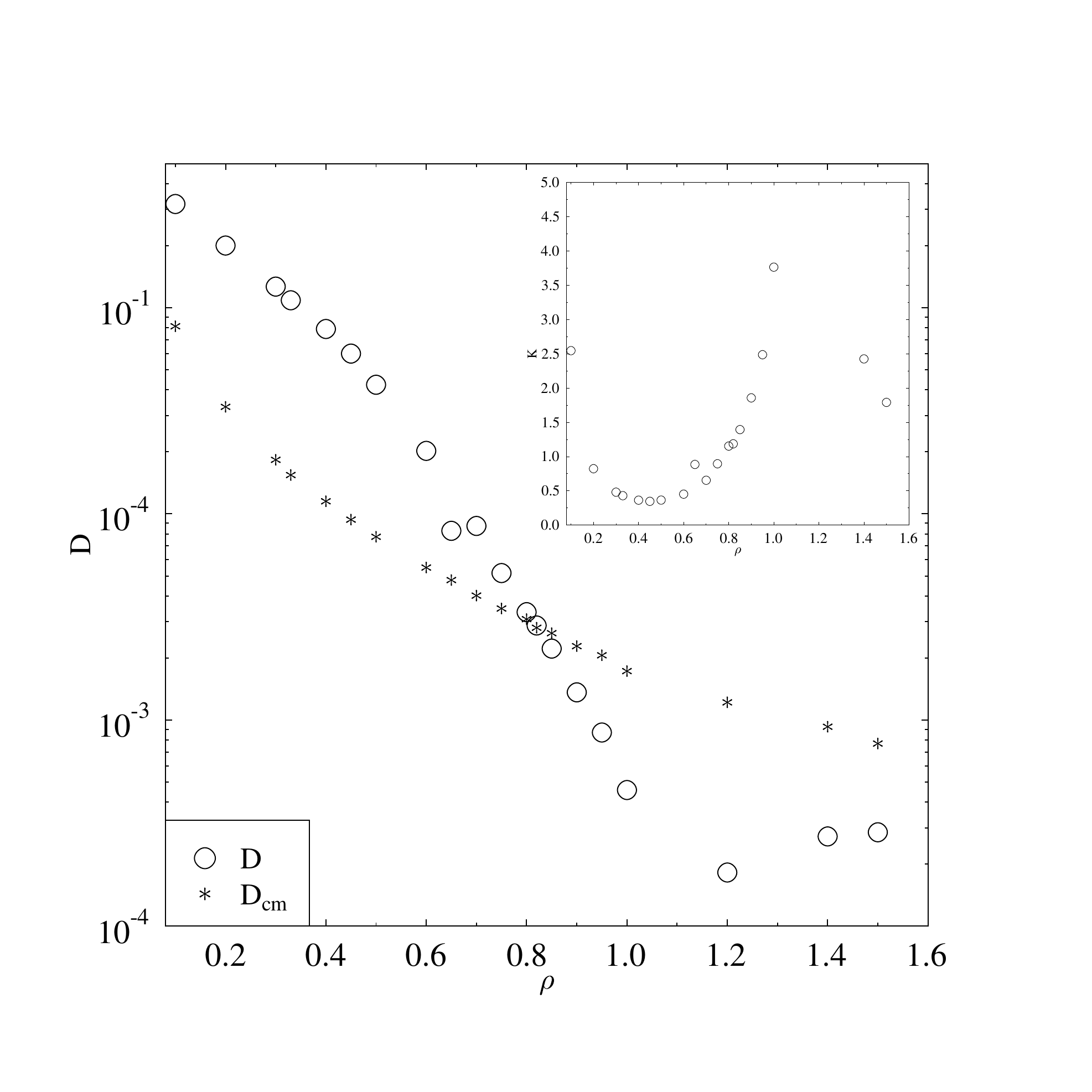}
	\end{center}
	\caption{The diffusion coefficient $D(\rho)$ and the
	compressibility $K(\rho)$ for $b=5$ and $L=100$.}
	\label{fig:diffusion}
\end{figure}

Here we are mainly interested in the possible signal of
the 'finite-size transition' via the compressibility,
which is shown in Fig.~\ref{fig:diffusion} in such a way
that $D$ has been obtained by the density-fluctuation
method and $D_{cm}$ from the center-of-mass mean-square
displacement,{\footnote{The coefficients
$D$ as obtained from density fluctuations and $D_{cm}$
as obtained from the center-of-mass mean-square displacement were
determined from fits over time windows of same length, which
above $\rho_c$ is considerably less than the time scale
of metastability. Therefore, $D$ and $D_{cm}$ represent an
average over two kinds of dynamics (cf.~Fig. 1).
We note also that they are sensitive to the topology of
the transition graph and most appropriate for the case
with symmetric jumps considered here.}} and 
from these we obtain $K$ via Eq.~(\ref{eq:KG}).
For this case, we have $\rho_c =1/3$
and $\rho_c(L)\approx 0.8$ (see Fig.~\ref{fig:scaling}).
No sharp signal of the transition is seen, cf.~a peak in 
susceptibility at a second-order phase transition,
but around $\rho_c(L)$ the thermodynamic contribution 
does reach its minimum and,
consequently, the diffusion coefficient decays much 
faster than the mobility as a function of the density.

\newpage
\section{\label{sec:conclusion}Conclusions} 

The finite size effects present in a zero-range process 
in the symmetric one-dimensional case were studied by 
using continuous-time Monte Carlo simulations. 
It was shown that well above the thermodynamic critical point 
both static and dynamic properties display fluid-like behavior 
up to a density $\rho_c(L)$, which was determined from the 
turning point of the $Z_1(\rho)/N$ curve or the maximum of the 
corresponding susceptibility. The finite-size analysis of 
$\rho_c(L)$ gave results consistent with the existing scaling theory. 
We then analyzed the behavior of various static and dynamics quantities 
around $\rho_c(L)$ to demonstrate its physical significance. 
A detailed theoretical analysis of the crossover and the 
persistence properties of the largest cluster would 
certainly be of interest.


\begin{acknowledgments}
This research has been supported by the Magnus Ehrnrooth Foundation. 
\end{acknowledgments}

\bibliography{station}

\end{document}